\documentclass[prl,twocolumn,aps,epsfig,prbbib]{revtex4}
\usepackage{graphicx}

\begin{document}

\title{Unconventional approaches to 
combine optical transparency with electrical conductivity}

\author{J.~E. Medvedeva}
\email[]{E-mail:juliaem@umr.edu}

\affiliation{Department of Physics, University of Missouri--Rolla, Rolla, MO 65409}


\begin{abstract}

Combination of electrical conductivity and optical transparency 
in the same material -- known to be a prerogative of only a few oxides 
of post-transition metals, such as In, Sn, Zn and Cd --
manifests itself in a distinctive band structure of the transparent conductor host. 
While the oxides of other elements with $s^2$ electronic configuration, 
for example, Mg, Ca, Sc and Al, also exhibit the desired optical and electronic 
features, they have not been considered as candidates for 
achieving good electrical conductivity because of the challenges of 
efficient carrier generation in these wide-bandgap materials.
Here we demonstrate that alternative approaches to the problem  
not only allow attaining the transport and optical properties which compete 
with those in currently utilized transparent conducting oxides (TCO),
but also significantly broaden the range of materials with a potential
of being developed into novel functional transparent conductors.
\end{abstract}

\maketitle

\begin{figure}
\includegraphics[width=4.0cm]{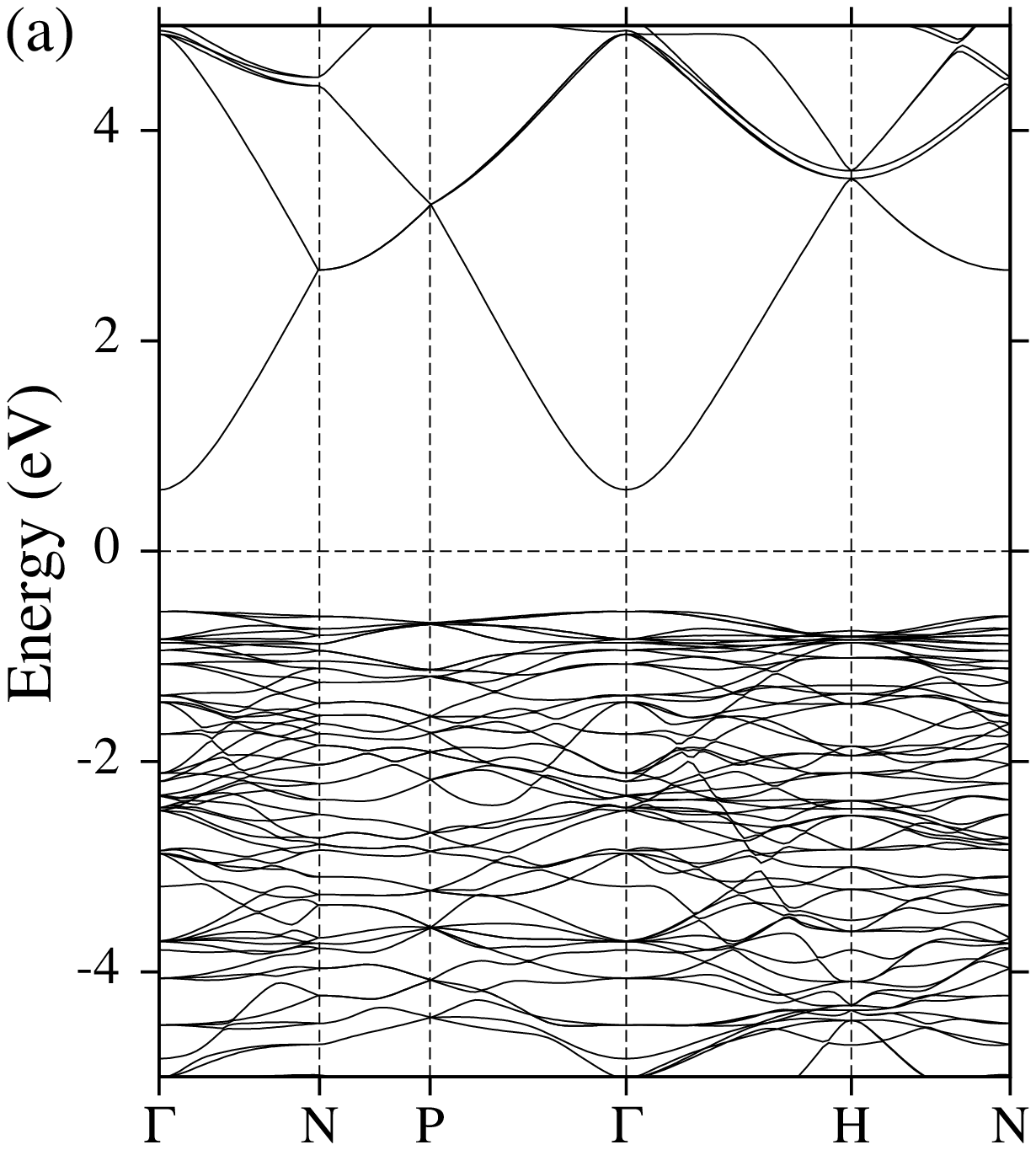}
\includegraphics[width=4.0cm]{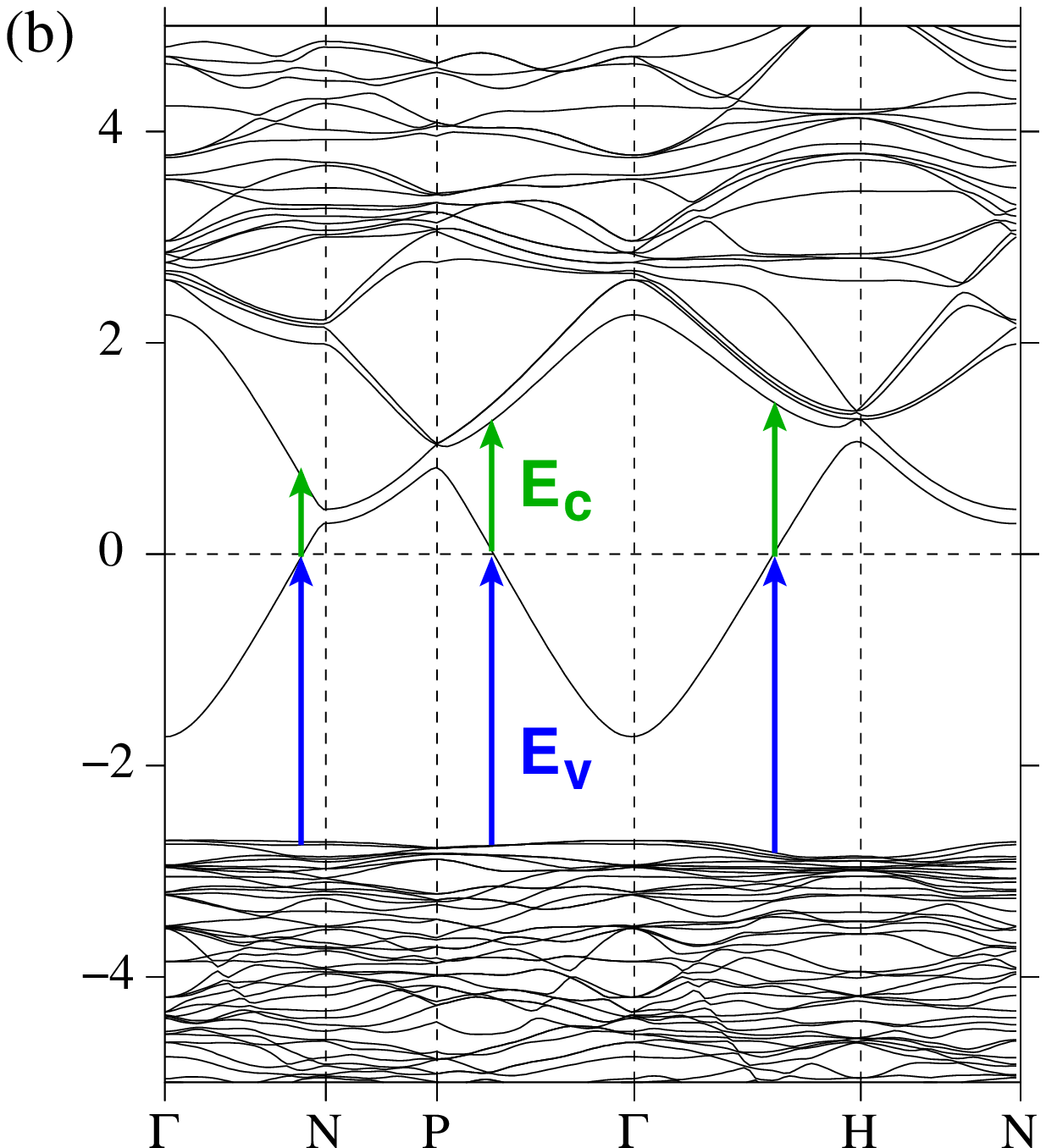}
\includegraphics[width=4.0cm]{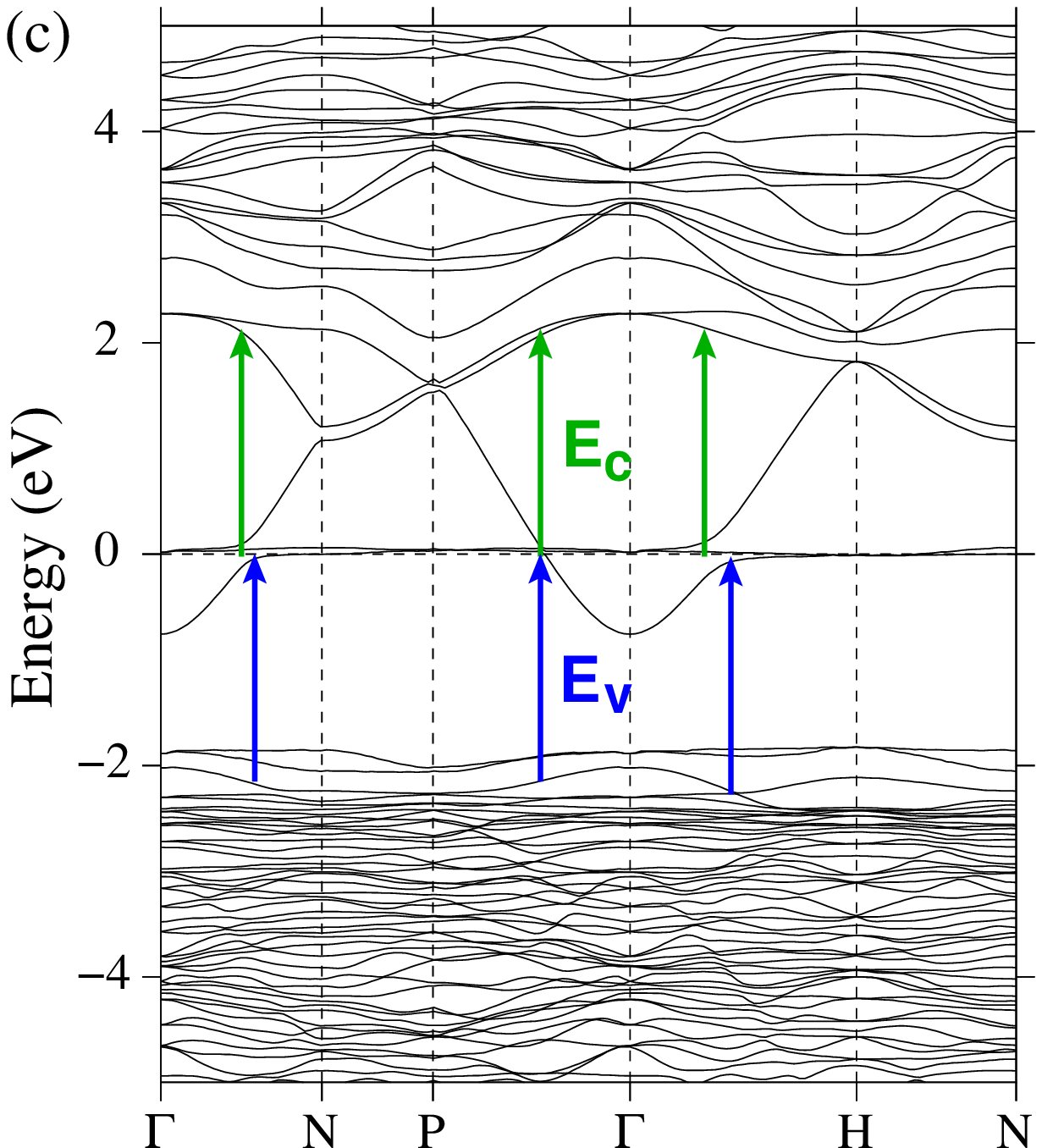}
\includegraphics[width=4.0cm]{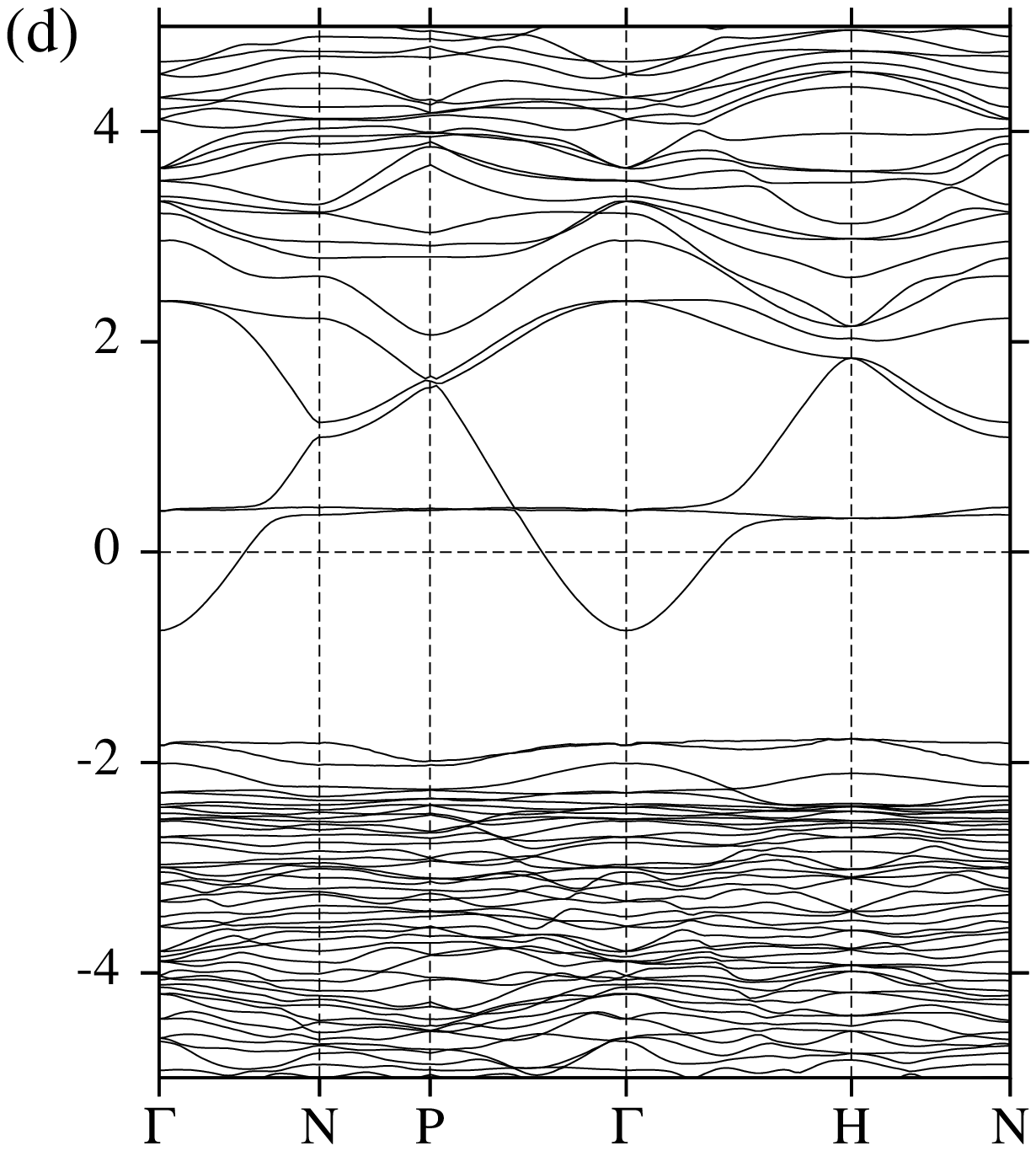}
\caption{Electronic band structure of pure (a), 6.25\% Sn-doped (b), and
6.25\% Mo-doped In$_2$O$_3$ for the majority (c) and the minority (d) 
spin channels.}
\label{imo}
\end{figure}

The key attribute of any conventional n-type TCO host is a highly dispersed 
single free-electron-like conduction band 
\cite{Freeman,Mryasov,Asahi,Woodward,EPL,JACS,JACS1,my-PRL}.
Upon proper doping, it provides both (i) high mobility of extra carriers
(electrons) due to their small effective mass, and
(ii) low optical absorption in the visible part of the spectrum 
due to high-energy inter-band transitions, e.g., Fig. \ref{imo}.
For the complete transparency in the visible range, the transitions 
from the valence band, E$_v$, and from the partially filled conduction 
band, E$_c$, should be larger than 3.1 eV, 
while the intra-band transitions as well as the plasma frequency
should be smaller than 1.8 eV.
The high energy dispersion also ensures a pronounced Fermi energy displacement, 
so-called Burstein-Moss (BM) shift, so that the optical transparency can be 
achieved in a material with a relatively small bandgap, for example, in CdO where 
the optical (direct) band gap is 2.3 eV.

Figures \ref{imo}(a) and \ref{imo}(b) illustrate the typical conduction band 
of a conventional n-type transparent conductor 
and how doping alters the electronic band structure of 
the TCO host affecting the optical transitions.
It is seen that upon introduction of extra carriers into the host,
a large BM shift which facilitates higher-energy transitions from the valence band (E$_v$), 
leads to a reduced energy of the transitions from 
the Fermi level up into the conduction band (E$_c$), i.e., 
E$_v$ and E$_c$ are interconnected \cite{second-hybr-gap}.
In other words, large carrier concentrations 
desired for a good conductivity, may result in an increase of the optical 
absorption because the E$_c$ transitions become smaller in energy. 
In addition, the transitions within the partially filled 
band as well as plasma frequency may lead to the absorption in the long-wavelength range.

The mutual exclusiveness of the optical transmittance and electrical 
conductivity (see Refs. \cite{EPL,Bellingham,MRS-Coutts}) makes it challenging 
to achieve the optimal performance in a transparent conductor.
Below we outline novel, unconventional ways to balance the optical and transport
properties and to improve one without making a sacrifice of the other.

\subsection{Magnetically mediated transparent conductors}

One of the possible routes to avoid compromising the optical transparency is 
to enhance conductivity 
via mobility of the carriers rather than their concentration \cite{MRS-Coutts}. 
Recently, the mobility with more than twice the value of the commercial 
Sn-doped indium oxide (ITO) was observed in Mo-doped In$_2$O$_3$ (IMO),
and it was shown that the conductivity can be significantly increased
with no changes in the spectral transmittance upon doping with Mo
\cite{Meng,Yoshida,Yoshida-effmass,Sun}.
Surprisingly, introduction of the transition metal Mo which donates
two more carriers per substitution compared to Sn,
does not lead to the expected increase of the optical absorption
or a decrease of the mobility due to the scattering on the localized 
Mo $d$-states.

\begin{table}
\begin{tabular}{l|cc|ccc|c} \hline
Complex & \, M \, & E$_g$(0) & k$_F^{[110]}$ & k$_F^{[111]}$ & k$_F^{[010]}$ & $\omega_p$ \\ \hline

Mo$^{\bullet\bullet\bullet}_{In(1)}$ & 1.85 & 1.38 & \, 0.152 \, & \, 0.148 \, & \, 0.157 \, & \, 1.63 \, \\

Mo$^{\bullet\bullet\bullet}_{In(2)}$ & 1.32 & 1.18 & 0.194 & 0.187 & 0.201 & 2.05 \\

[Mo$^{\bullet\bullet\bullet}_{In}$O$^{''}_i$]$^{\bullet}$ & 0.50 & 1.26 & 0.125 & 0.123 & 0.132 & 1.27 \\ \hline

Sn$^{\bullet}_{In(1)}$ & --- & 0.98 & 0.201 & 0.203 & 0.205 & 2.29 \\

In$_2$O$_3$+e$^{'}$ & --- & 1.16 & 0.206 & 0.204 & 0.213 & 2.38 \\ \hline

\end{tabular}
\caption{Calculated magnetic moments on the Mo atoms, M, in $\mu_B$;
the fundamental band gap values E$_g$(0), in eV; the Fermi wave vectors k$_F$, in atomic
units; and the plasma frequency $\omega_p$, in eV,
for the different substitutional complexes with 6.25\% Mo doping level.
Calculated values for pure (rigid-band model) and 6.25\% Sn-doped In$_2$O$_3$ are given for comparison.}
\label{table}
\end{table}

Our electronic band structure investigations of IMO revealed
\cite{my-PRL} that the {\it magnetic interactions} 
which have never been considered to play a role in combining optical 
transparency with electrical conductivity, ensure both high carrier mobility 
and low optical absorption in the visible range.
As one can see from Figs. \ref{imo}(c) and \ref{imo}(d), 
strong exchange interactions split 
the Mo $d$-states located in the vicinity of the Fermi level.
These $d$-states are resonant states, while the conductivity is due 
to the delocalized In $s$-states which form the highly dispersed 
free-electron-like conduction band. In other words, the free carriers 
in the system flow in a background of the Mo defects
which serve as strong scattering centers.
Because of the exchange splitting of the Mo $d$-states, the carriers of 
one spin is affected by only a half of the scattering centers, i.e., 
only by the Mo
$d$-states of the same spin. Therefore, the concentration of the Mo scattering
centers is effectively lowered by half compared to the Mo doping level.

Figs. \ref{imo}(b) and \ref{imo}(c,d) show that the BM shift 
is less pronounced in the IMO case -- despite the fact that Mo$^{6+}$ donates 
two extra carriers as compared to Sn$^{4+}$ at the same doping level.
Such a low sensitivity to doping appears from the resonant 
Mo $d$-states located at the Fermi level that facilitates the $d$-band 
filling (pinning) and thus hinders further displacement of the Fermi level 
deep into the conduction band.
Smaller BM shift in IMO leads to the following advantageous features 
to be compared to those of ITO:

(i) Smaller increase in the effective mass is expected upon Mo doping.
In addition, the resonant Mo $d$-states do not hybridize with the $s$-states of indium and 
so do not affect the dispersion of the conduction band.
Therefore, the effective mass remains similar to the one of pure indium oxide.
This is borne out in experimental observations \cite{Yoshida-effmass} showing
that the effective mass does not vary with doping (up to 12 \% of Mo)
and/or carrier concentration.

(ii) Larger (in energy) optical transitions from the partially occupied band 
(cf., Figs. \ref{imo}(b) and \ref{imo}(c,d)) along with the fact that 
transitions from $d$- to $s$-states
are forbidden ensure lower short-wavelength optical absorption.

(iii) The calculated plasma frequency, $\omega_p$, in IMO is below the visible range and
significantly smaller than that of ITO (Table \ref{table}).
This finding suggests a possibility to introduce larger carrier concentrations
without sacrificing the optical transmittance in the long wavelength range.

(iv) Smaller BM shift does not lead to the appearance of the intense inter-band 
transitions from the valence band, E$_v$, in the visible range due to the large 
optical band gap in pure indium oxide, namely, 3.6 eV \cite{Hamberg}.
Furthermore, in contrast to ITO where the bandgap narrowing
has been demonstrated both experimentally \cite{Hamberg}
and theoretically \cite{Mryasov}, doping with Mo shows an opposite 
(beneficial) trend: the fundamental band gap increases upon introduction 
of Mo, Table \ref{table}, because
the asymmetric $d$-orbitals of Mo rotate the $p$-orbitals 
of the neighboring oxygen atoms leading to an increase of the overlap 
between the latter and the In $s$-states.

It is important to note that the optical and transport properties in IMO 
are sensitive to specific growth conditions, namely, the ambient oxygen pressure.
It is found \cite{my-PRL} that 
an increased oxygen content facilitates the formation of the oxygen 
compensated complexes which reduces the number of free carriers -- 
from 3 to 1 per 
Mo substitution -- but, at the same time, improves the carrier mobility
due to smaller ionized impurity scattering and hence longer relaxation times.
On the other hand, the interstitial oxygen significantly supresses 
the magnetic interactions, Table \ref{table}, 
which should be strong enough to split the transition metal $d$-states 
in order to provide good conductivity in one (or both) spin channels.

Thus, the transition metal dopants can be highly beneficial in providing the transport 
and optical properties which compete with those of commercially utilized ITO. 
Similar behavior is expected upon doping with other transition metal 
elements and other hosts -- provided that the magnetic interactions are small enough 
to keep the $d^{\uparrow}$-$d^{\downarrow}$ transitions out of the visible range.

\subsection{Multicomponent TCO with layered structure}

Complex transparent conductors consisting of structurally and/or 
chemically distinct layers, 
such as InGaO$_3$(ZnO)$_m$, $m$=integer, offer a way to increase conductivity 
by spatially separating 
the carrier donors (traditionally, oxygen vacancies or aliovalent substitutional dopants) 
and conducting layers which transfer the carriers effectively, 
i.e., without charge scattering on the impurities \cite{Li-implant,exp,cluster}. 

The homologous series InGaO$_3$(ZnO)$_m$ and In$_2$O$_3$(ZnO)$_m$, 
also known for their promising thermoelectric properties \cite{thermoelectr}, 
have been extensively studied experimentally 
\cite{Freeman,exp,cluster,Hiramatsu,PhilMag,Science,Nature,a-mass}.
In these materials, octahedrally coordinated In layers 
alternate with $(m+1)$ layers of oxygen tetrahedrons around Zn (and Ga)
\cite{str,str-all,family}. 
Because the octahedral oxygen coordination of cations was 
long believed to be essential for a good transparent conductor
\cite{Freeman,Woodward,Li-implant,Shannon,spinel-review,KawazoeMRS,Mason-review},
it has been suggested that the charge is transfered within the InO$_{1.5}$ layers
while Ga and Zn atoms were proposed as candidates 
for efficient substitutional doping \cite{Freeman,exp,cluster}.

\begin{figure}
\includegraphics[width=4.0cm]{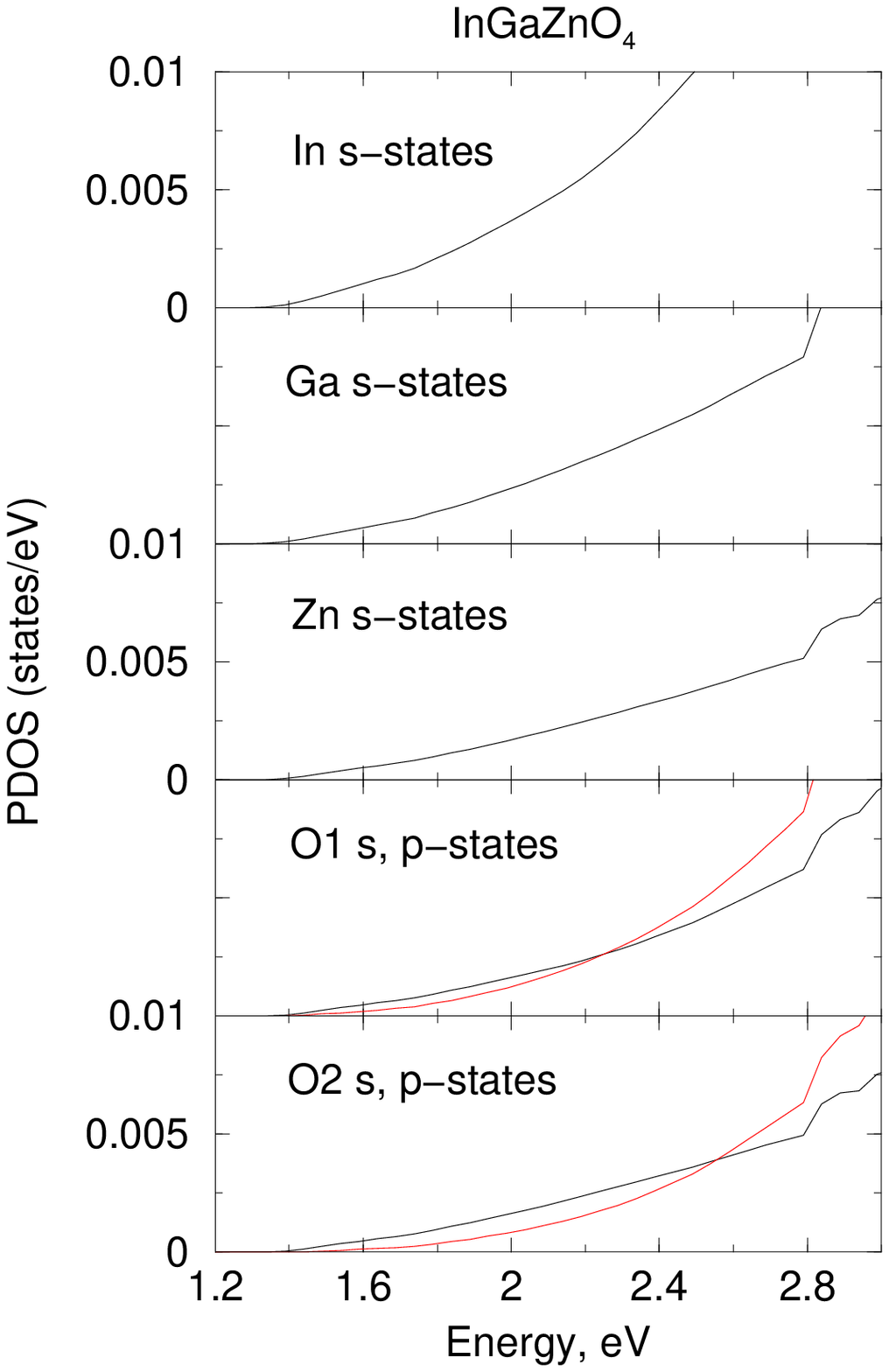}
\includegraphics[width=4.0cm]{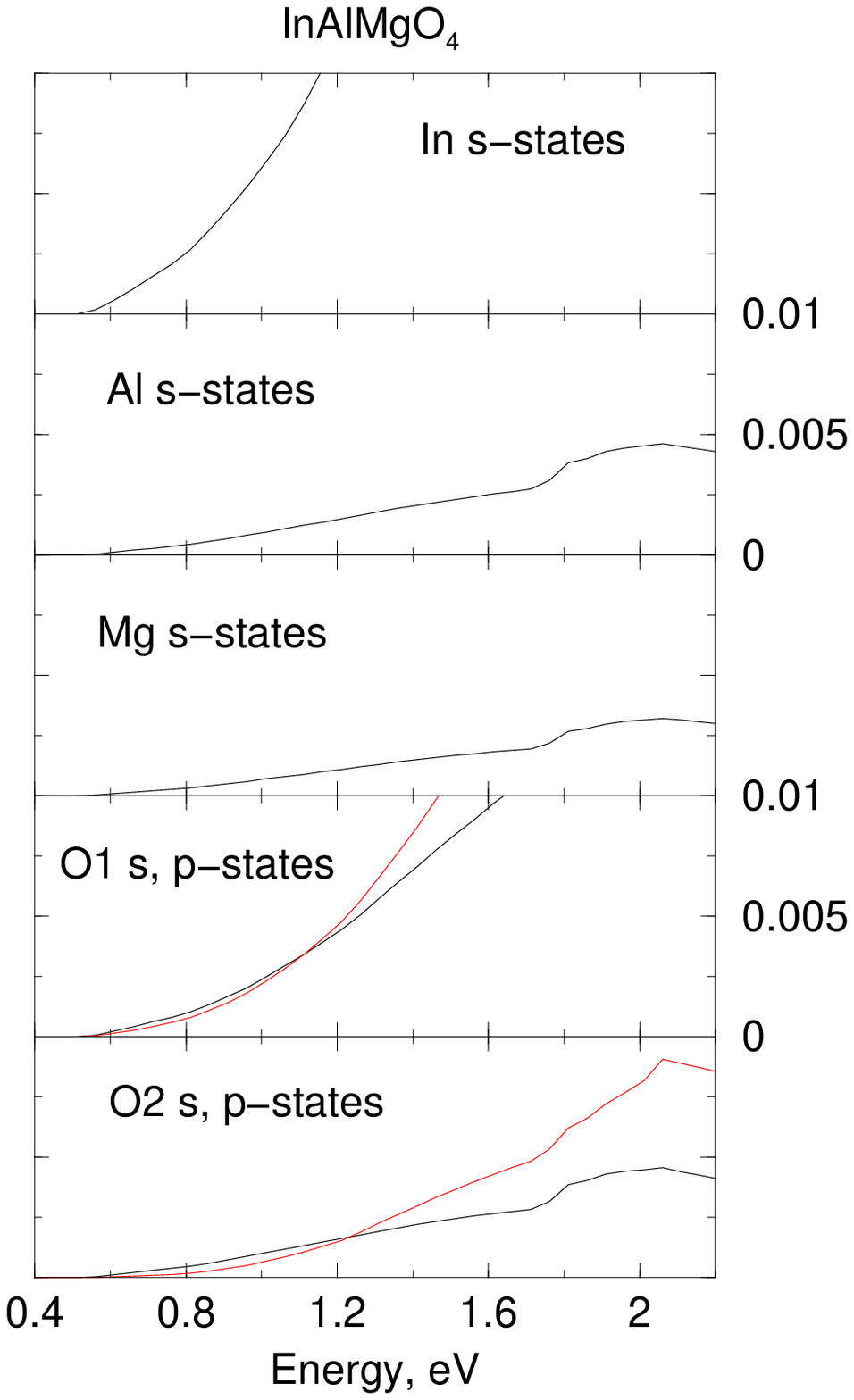}
\caption{Partial density of states for InGaZnO$_4$ and InAlMgO$_4$.
}
\label{contrib}
\end{figure}

However, accurate electronic band structure investigations for 
InGaZnO$_4$, $m$=1, showed \cite{JEM} that the atoms from both 
InO$_{1.5}$ and GaZnO$_{2.5}$ layers 
give comparable contributions to the conduction band, Fig. \ref{contrib}(a). 
This resulted in a three-dimensional distribution of the charge density:
the interatomic (or ``background'') electron density
is similar in and across the [0001] layers.
The isotropy of the electronic properties in this layered compound
manifests itself in the electron effective masses being nearly 
the same in all crystallographic directions (Table \ref{table-mass}).

\begin{table}
\begin{tabular}{l|cc|c|ccc|cc} \hline
Compound    & N$_1$ & N$_2$ & E$_g$(0) & m$_{[100]}$ & m$_{[010]}$ & m$_{[001]}$ & m$_{ab}$ & m$_z$ \\ \hline
InGaZnO$_4$ & 48\% & 52\%  &  1.30  & 0.23 & 0.22 & 0.20 & 0.23 & 0.23 \\
InGaMgO$_4$ & 58\% & 42\%  &  2.15  & 0.27 & 0.27 & 0.24 & 0.28 & 0.29 \\
ScGaZnO$_4$ & 26\% & 74\%  &  2.48  & 0.33 & 0.33 & 0.34 & 0.33 & 0.53 \\
InAlMgO$_4$ & 72\% & 28\%  &  2.78  & 0.32 & 0.31 & 0.35 & 0.31 & 0.34 \\ \hline
\end{tabular}
\caption{Net contributions to the conduction band at the $\Gamma$ point 
from the states of the atoms that belong to 
the In(Sc)O$_{1.5}$, N$_1$, or Ga(Al)Zn(Mg)O$_{2.5}$, N$_2$, layers, in per cent; 
the LDA fundamental bandgap values E$_g$(0), in eV; 
the electron effective masses m, in $m_e$, along the specified crystallographic directions; 
and the components of the electron effective-mass tensor, $m_{a,b}$ and $m_z$, calculated
via simple averaging of those of the corresponding single-cation oxides.}
\label{table-mass}
\end{table}

Most strikingly, we found that the effective mass remains isotropic when 
the cation(s) in InGaZnO$_4$ are replaced by other elements 
with $s^2$ electronic configuration, for example, Sc, Al and/or Mg.
This finding may seem to be counterintuitive, since the $s$-states
of Sc, Al and Mg are expected to be located deeper in the conduction band
due to significantly larger band gaps in Sc$_2$O$_3$, Al$_2$O$_3$ and MgO
as compared to those in In$_2$O$_3$, Ga$_2$O$_3$ and ZnO.
Analysis of the partial density of states shows that 
although the contributions from the Sc, Al and Mg atoms 
to the bottom of the conduction band are notably reduced, cf., Fig. \ref{contrib} 
and Table \ref{table-mass}, the states of these atoms are still available 
for the electron transport. 
Consequently, the interatomic charge density distribution is
three-dimensional for all these layered multi-cation oxides -- in accord with 
the isotropic electron effective mass.
Moreover, we found \cite{JEM} that the electron effective mass in these complex 
materials can be predicted via simple averaging over those of the corresponding
single-cation oxides (Table \ref{table-mass}).

It is important to stress that the isotropic character of the 
{\it intrinsic} transport properties in the TCO hosts 
with layered structure may not be maintained when 
extra carriers are introduced. 
Different valence states (In$^{3+}$ and Ga$^{3+}$ vs Zn$^{2+}$) 
and oxygen coordination (octahedral for In vs tetrahedral for Ga and Zn) 
are likely to result in preferential (non-uniform) arrangement of 
aliovalent substitutional dopants or oxygen vacancies.
We believe that the observed anisotropic conductivity 
\cite{thermoelectr,Hiramatsu} as well as its dependence  
on the octahedral site density \cite{Freeman,Mason-review}
in the layered TCO's is a manifestation of a specific carrier generation mechanism.
While proper doping can help make either or both structurally distinct 
layers conducting, leading to a highly anisotropic or three-dimensional
electron mobility, respectively, amorphous complex oxides \cite{PhilMag,Nature,a-mass}
readily offer a way to maintain isotropic transport properties. 

Thus, we believe that other cations with $s^2$ electronic configuration,
beyond the traditional In, Sn, Zn and Cd, 
can be effectively incorporated into novel complex multicomponent TCO hosts --
such as the layered materials decribed above, ordered ternary oxides 
\cite{Mason-review,Minami,Minami05}, solid solutions \cite{Freeman} 
as well as their amorphous counterparts \cite{Minami05,PhilMag}, 
important for flexible electronics technologies \cite{Nature}.
Significantly, the sensitivity of the bandgap value to the composition 
of a multicomponent oxide (Table \ref{table-mass}) offers a possibility to
manipulate the optical properties as well as the band offsets 
(work functions) via proper composition of an application-specific TCO. 

Finally, it should be mentioned, that the efficient doping of the wide-bandgap oxides
is known to be a challenge \cite{Neumark,VandeWalle,Zunger}. 
Alternative carrier generation mechanisms, for example, magnetic dopants 
discussed above, introduction of hydrogen \cite{ValleNature}
or ultraviolet irradiation in nanoporous calcium aluminate 
\cite{EPL,mayenite,PRLmay}, are being actively sought
and have already yielded promising results -- as outlined in the following section.



\subsection{Novel UV-activated transparent conductors}

\begin{figure}
\includegraphics[width=4cm]{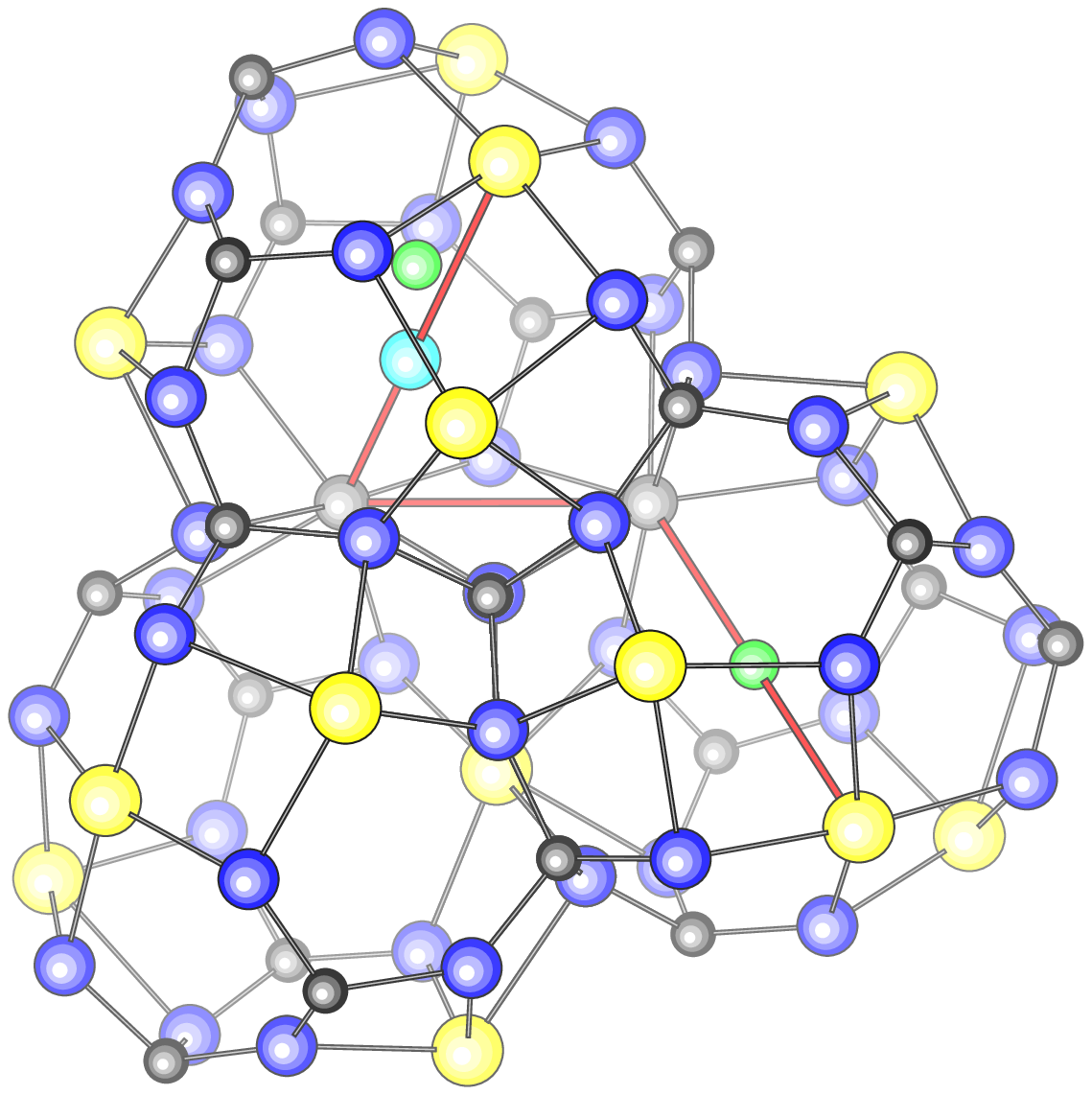}
\includegraphics[width=4cm]{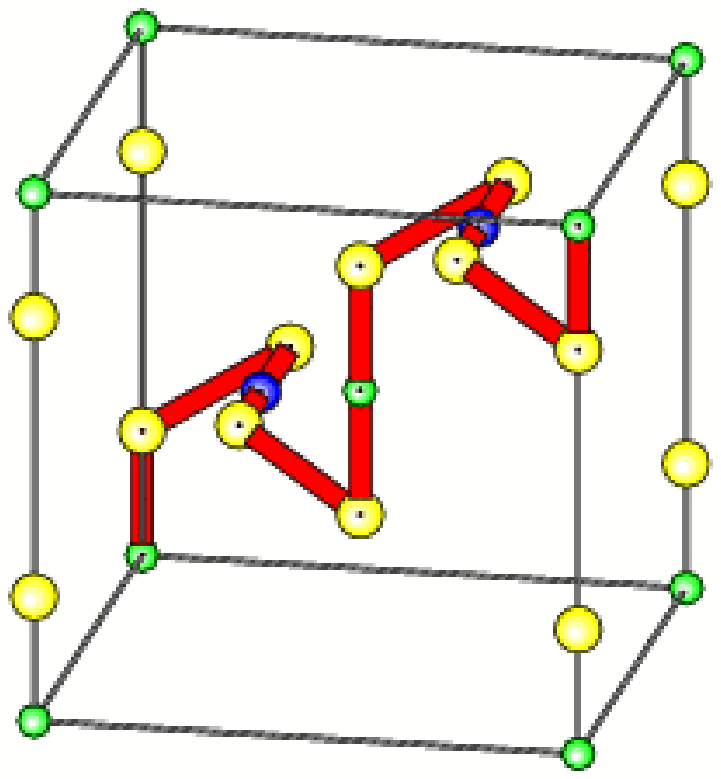}
\caption{Crystal structure of H-doped 12CaO$\cdot$7Al$_2$O$_3$. On the left, 
only three of the 12 cages in the unit cell are shown. The cube (right)
represent the unit cell. Only the atoms that participate in hopping transport 
are shown: Ca (yellow), OH$^-$ (blue) and H$^-$ (green spheres). Red line
represent the electron hopping path.
}
\label{3cage}
\end{figure}

Cage-structured insulating calcium-aluminum oxide, 12CaO$\cdot$7Al$_2$O$_3$, or mayenite, 
differs essentially from the conventional TCO's not only by its chemical and structural 
properties but also by the carrier generation mechanism: a persistent conductivity 
(with a ten-order of magnitude change)
has been achieved upon doping with hydrogen followed by UV irradiation \cite{mayenite,PRLmay}. 

\begin{figure}
\includegraphics[width=2.7cm]{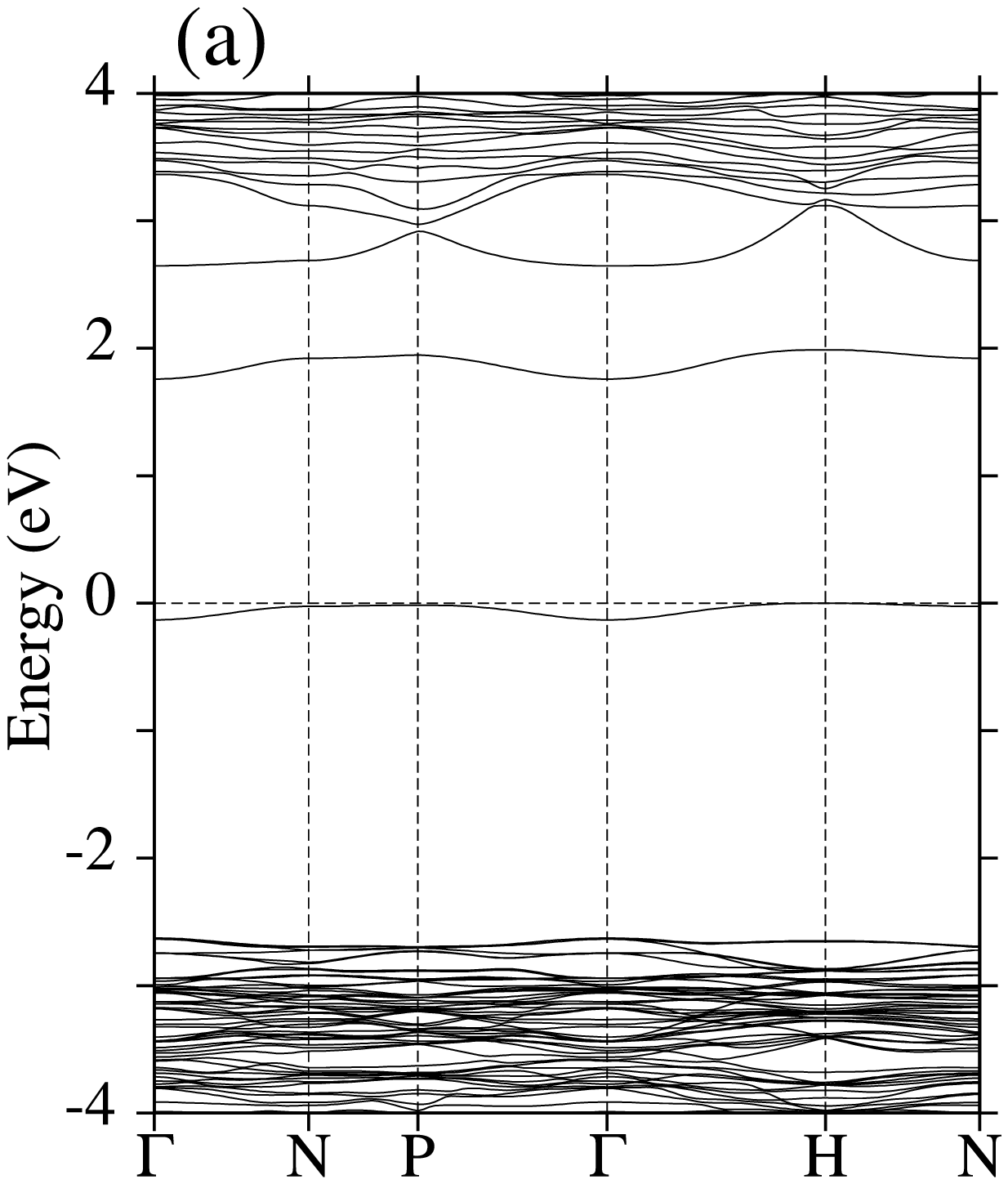}
\includegraphics[width=2.7cm]{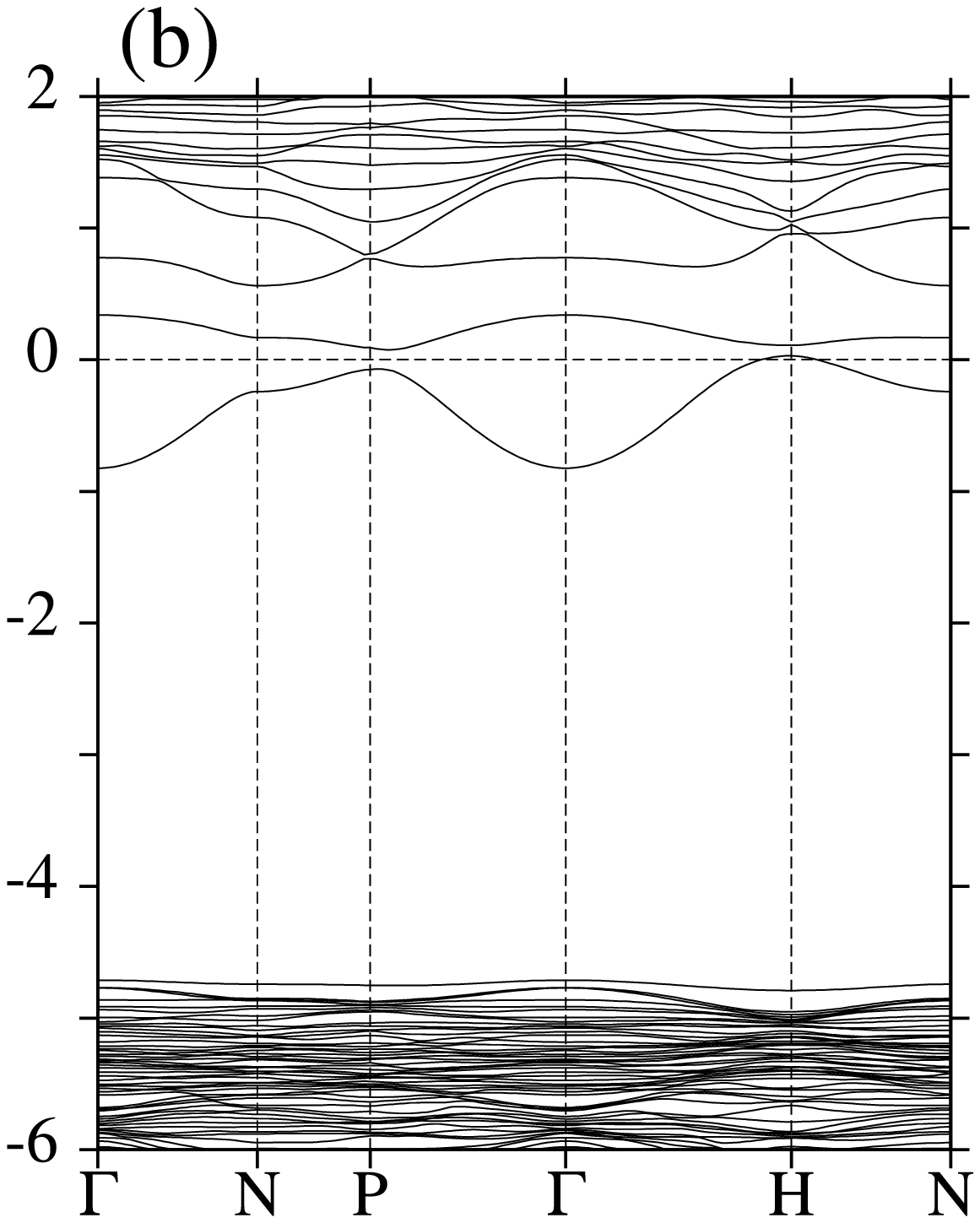}
\includegraphics[width=2.7cm]{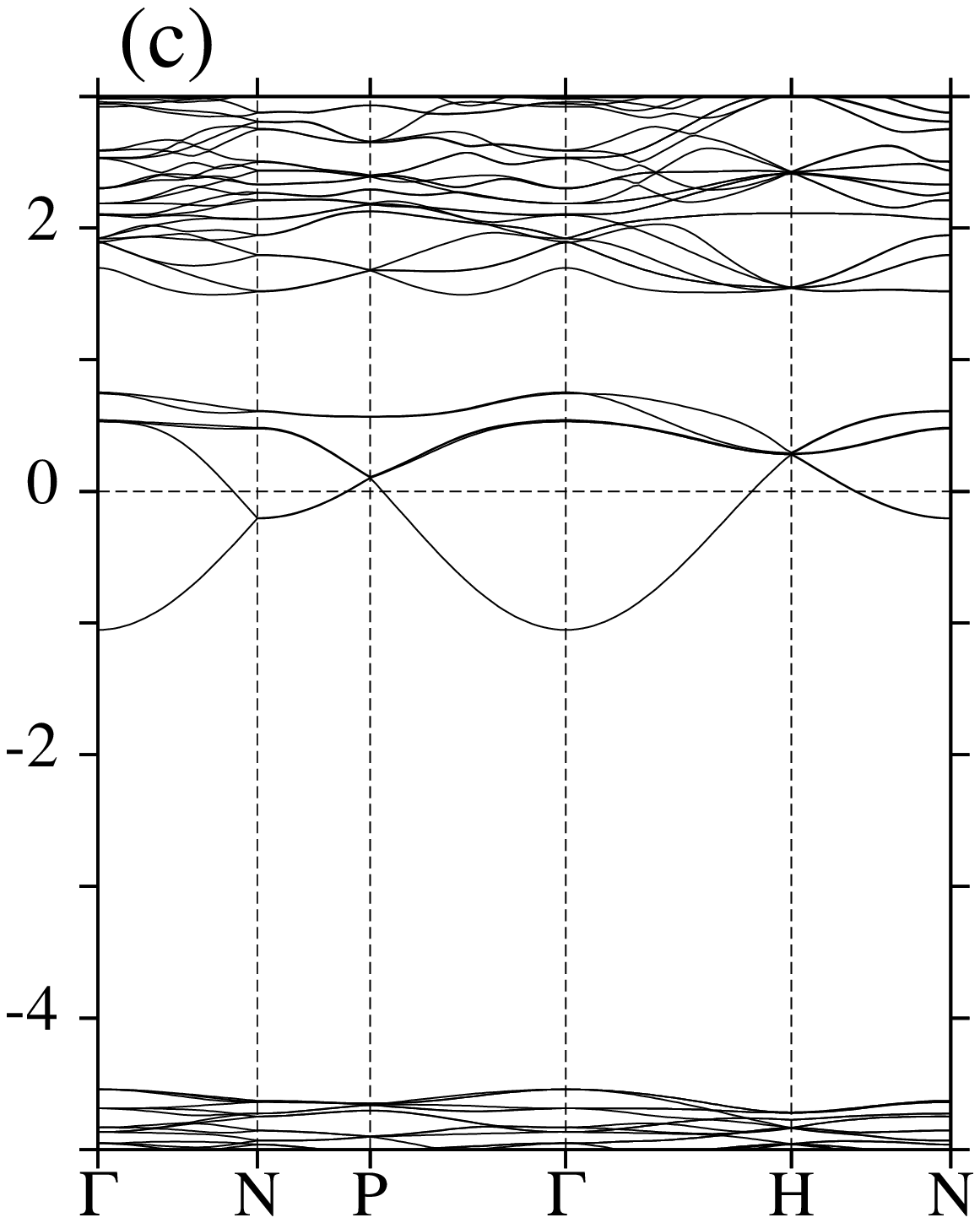}
\caption{Electronic band structure of (a) H-doped and (b) H-doped UV-irradiated
Ca$_{12}$Al$_{14}$O$_{33}$ and (c) [Ca$_{12}$Al$_{14}$O$_{32}$]$^{2+}$(2e$^-$).}
\label{may-bands}
\end{figure}

Mayenite belongs to the CaO-Al$_2$O$_3$ family of Portland cements which are known 
for their superior refractory properties. The unique structural features of mayenite, 
Fig. \ref{3cage}, namely, the encaged ``excess'' oxygen ions, allow incorporation of hydrogen
according to the chemical reaction: O$^{2-}$(cage) + H$_2$(atm.) $\rightarrow$ OH$^-$(cage) + 
H$^-$(another cage).
While the H-doped mayenite remains insulating (Fig. \ref{may-bands}(a)), the conductivity 
results from the electrons excited by UV irradiation off the H$^-$ ions into the conduction 
band formed from Ca $d$-states. The charge transport  occurs by electron hopping 
through the encaged ``defects'' -- the H$^0$ and OH$^-$ located 
inside the large (more than 5.6 \AA \, in diameter) structural cavities.
Understanding of the conduction mechanism on the microscopic level \cite{PRLmay,electride}
resulted in prediction of ways to control the conductivity by targeting the particular 
atoms that participate in the hopping. 
These predictions have been confirmed experimentally \cite{PRLmay,Bertoni}.

The low conductivity in H-doped UV-irradiated mayenite ($\sim$1 S/cm, Ref. \cite{mayenite,PRLmay}) 
was attributed to the strong Coulomb interactions between the UV released electrons 
which migrate along a narrow conducting channel -- the hopping path, Fig. \ref{3cage}.
Alleviation of the electronic repulsion \cite{electride} resulted in 
the observed \cite{Matsuishi2003} 100-fold enhancement of the conductivity 
in the mayenite-based oxide, [Ca$_{12}$Al$_{14}$O$_{32}$]$^{2+}$(2e$^-$),
although the carrier concentration was only two times larger than that
in the H-doped UV-irradiated Ca$_{12}$Al$_{14}$O$_{33}$.
The improved conductivity, however, came at the cost of greatly increased
absorption \cite{electride,Matsuishi2003} due to an increased density of states 
at the Fermi level, making this oxide unsuitable 
for practical use as a transparent conducting material.

\begin{figure}
\centerline{
\includegraphics[height=5.8cm]{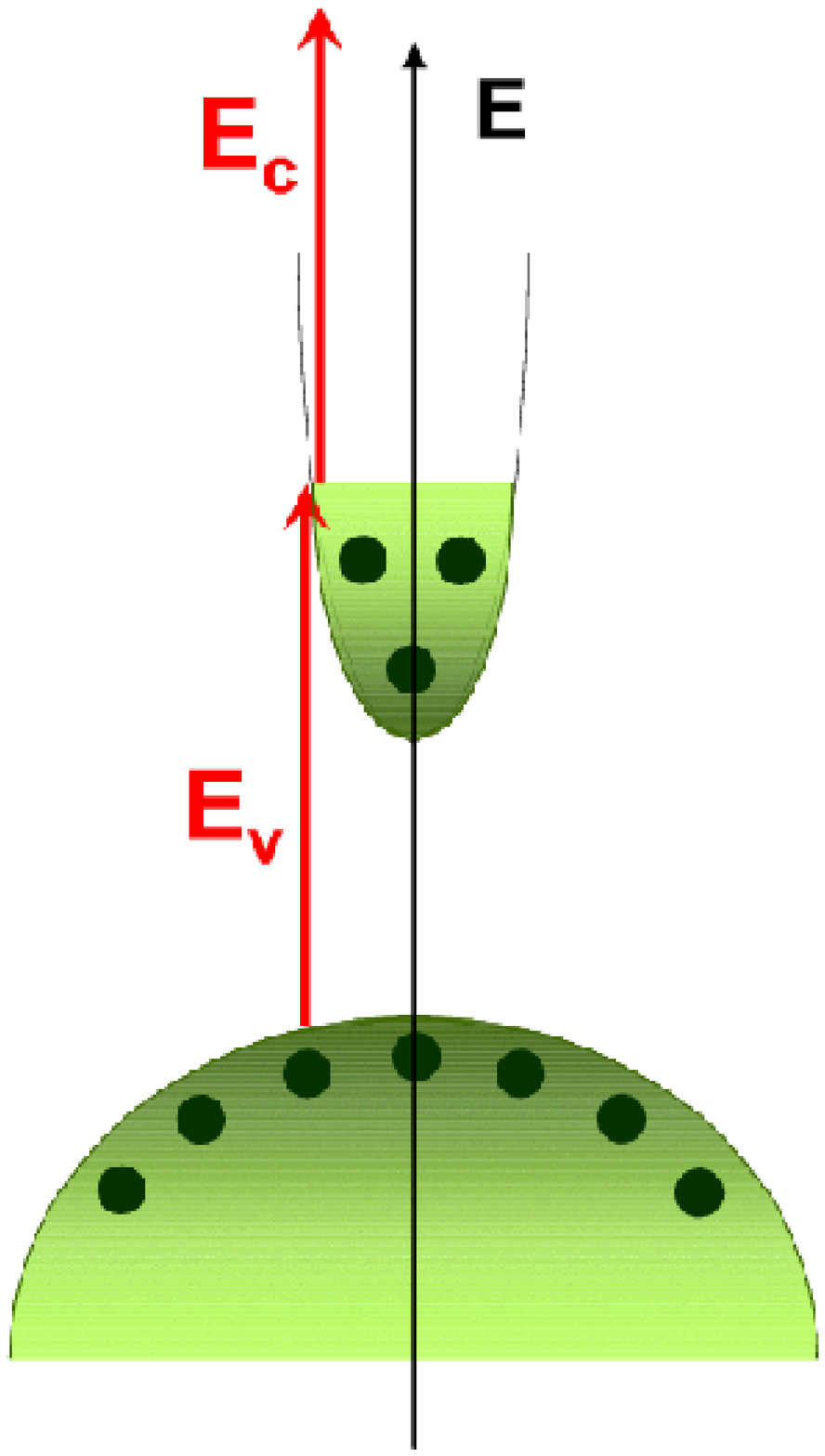}
\includegraphics[height=6cm]{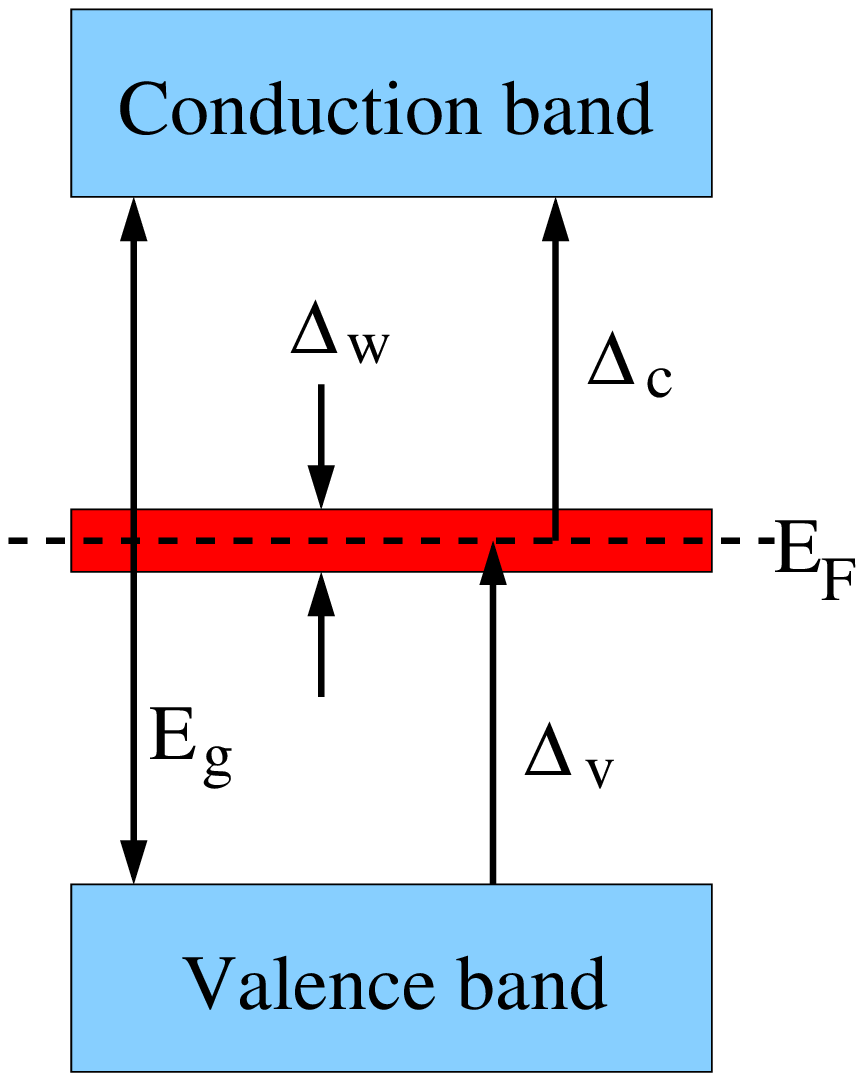}
}
\caption{Schematic band structure of conventional (left) and ``ideal'' (right)
transparent conductor.}
\label{scheme}
\end{figure}

Despite the failure to combine effectively the optical transparency and 
with useful electrical conductivity in Ca$_{12}$Al$_{14}$O$_{33}$,
the band structure analysis of the mayenite-based oxides suggests 
that these materials belong to a conceptually 
new class of transparent conductors where a significant correlation between 
their structural peculiarities and electronic and optical properties allows 
achieving good conductivity without compromising their optical properties. 
In striking contrast to the conventional TCO's, where there is a trade-off
between optical absorption and conductivity, as discussed above, 
nanoporous materials allow a possibility to combine 100\% optical 
transparency with high electrical conductivity.
The schematic band structure of such an ``ideal'' TCO is shown
in Fig. \ref{scheme}. Introduction of a {\it deep} impurity band 
in the bandgap of an insulating 
material would help to keep intense interband transitions (from the valence
band to the impurity band and from the impurity band to the conduction band)
above the visible range. 
This requires the band gap of a host material to be more than 6.2 eV. 
Furthermore, the impurity band should be narrow enough (less than 1.8 eV) 
to keep intraband transitions (as well as the plasma frequency) 
below the visible range.

In order to achieve high conductivity, the concentration of impurities should be 
large enough so that their electronic wavefunctions overlap and form
an impurity {\it band}. The formation of the band would lead to a high 
carrier mobility due to the extended nature of these states 
resulting in a relatively low scattering. For this, a material with 
a close-packed structure should {\it not} be used, 
because large concentration of impurities would result in
(i) an increase of ionized impurity scattering which limits 
electron transport; and (ii) a large 
relaxation of the host material, affecting its electronic structure and, 
most likely, decreasing the desired optical transparency.
Therefore, an introduction of a deep impurity band into a wide-band insulator
with a close-packed structure would make the material neither conducting
nor transparent.
Alternatively, materials with a nanoporous structure may offer a way to incorporate a
large concentration of impurities without any significant changes in
the band structure of the host material. 
Zeolites have been proposed \cite{EPL} as potential candidates for the ``ideal'' TCO's 
because they possess the desired structural and optical features, i.e., 
spacious interconnected pores and large bandgaps, and also exhibit the ability to trap 
functional ``guest'' atoms inside the nanometersized cavities which 
would govern the transport properties of the material.

\vspace{0.5cm}

Thus, understanding the principles of the conventional transparent 
conductors provide a solid base for further search of novel TCO host materials
as well as efficient carrier generation mechanisms.
Ab-initio density-functional band structure investigations \cite{methods} 
are valuable not only in providing
a thorough insight into the TCO basics but also in predicting hidden capabilities 
of the materials beyond the traditionally employed.




\begin{thebibliography}{99}
\bibitem{second-hybr-gap}
We note that in Sn-doped In$_2$O$_3$ (ITO) the second, or
so-called hybridization gap which 
was argued to be one of the reasons of low optical absorption in ITO 
\cite{Freeman,Mryasov} closes upon structural relaxation,
i.e., when the Sn-O and In-O distances are optimized.
\bibitem{methods}
We employed highly-precise all-electron full-potential linearized augmented plane 
wave (FLAPW) method \cite{FLAPW,FLAPW1} 
with the local density approximation for the electronic band structure
investigations of Mo-doped In$_2$O$_3$, multicomponent $XY_2$O$_4$ compounds, 
H-doped Ca$_{12}$Al$_{14}$O$_{33}$ and oxygen deficient 
[Ca$_{12}$Al$_{14}$O$_{32}$]$^{2+}$(2e$^-$).
Each structure has been optimized via the total energy and atomic
forces minimization. During the optimization, all atoms were allowed to move 
along $x$, $y$, and $z$ directions while the lattice parameters
were fixed at the corresponding experimental values.
The UV-irradiated H-bearing Ca$_{12}$Al$_{14}$O$_{33}$ was investigated within
the linear muffin-tin orbital method (LMTO) in the atomic sphere approximation 
\cite{LMTO}.
\end{thebibliography}
\end{document}